\documentclass[a4paper,11pt]{article}
\pdfoutput=1 

\usepackage{jinstpub} 
\usepackage{amsmath}
\usepackage{float}
\title{Improving radiation hardness in space-based Charge-Coupled Devices: An experimental validation of a new pre-fabrication modelling technique}

\author[a]{S.~Parsons,}
\author[a,1]{D.~Hall, \note{Corresponding author.}}
\author[a]{T.~Buggey,}
\author[a]{A.~Holland,}
\author[b]{P.~Verhoeve}

\affiliation[a]{Centre of Electronic Imaging, Open University, \\Kents Hill, Milton Keynes, UK}
\affiliation[b]{ESTEC, Directorate of Science of the European Space Agency,\\Keplerlaan 1, Noordwijk, The Netherlands}

\emailAdd{david.hall@open.ac.uk}

\abstract{

The soft X-ray imager (SXI) on the SMILE mission uses two large 4510x4510 back illuminated CCD370s to detect X-rays in the 0.2-2~keV range. These devices take heritage from the optical imaging PLATO mission CCD270s and have been optimised for low energy signals by including a parallel supplementary buried channel (SBC) which should reduce the volume of the charge cloud and thereby reduce the number of traps it interacts with as it is transferred through the CCD.

The charge transfer performance improvement between the CCD270 and the CCD370 has been simulated using a combination of Silvaco and Matlab models to predict its characteristics pre-fabrication over a 10$^{2}$-10$^{5}$ electron signal range. Trap pumping measurements have been taken on both devices to count the number of traps present and hence calculate the mean amount of traps that exist per pixel across the range of signal levels. The trap pumping results are used to calculate a charge transfer performance improvement that  shows good agreement with the simulated values, especially in the SXI science band.

These results bring added confidence to the early performance modelling of the SMILE SXI instrument and is a good indicator that the simulations are accurate enough to be used to model devices with more advanced geometries such as an SBC and can be used in future CCD missions, where radiation-hardness and hence good charge transfer characteristics are key.
}

\keywords{CCD, Space instrumentation, radiation damage, X-ray detectors, SMILE}

\begin{document}
\maketitle
\flushbottom

\section{Introduction}
\label{sec:intro}
The Solar wind Magnetosphere Ionosphere Link Explorer (SMILE) mission is a joint venture between the European Space Agency (ESA) and the Chinese Academy of Sciences (CAS)~\cite{RaabSMILE}. The aim of SMILE is to develop the understanding of the Sun-Earth connection, for the first time on a global scale, by observing the solar wind interacting with the magnetosphere. This will be done through a combination of the Soft X-ray Imager (SXI), the Ultra-Violet Imager (UVI), the Light Ion Analyser (LIA) and the Magnetometer (MAG) instruments. The SXI will observe emission from the solar wind charge exchange (SWCX) process which generates low energy X-rays in the 0.2-2.0~keV energy range~\cite{DennerlChargeExchange}~\cite{WalshSXIscience}. SMILE will have a highly elliptical polar orbit and throughout its 3-year mission will receive radiation damage from trapped and solar protons~\cite{HollandDamage}. Combining this with the small signals to be detected and the direct path for any radiation through the micro pore optics (MPOs) means that understanding how the performance of the CCD370s within the SXI changes through their lifetime is essential for assessing whether the science goals will be achieved.  

The SMILE SXI will have two large CCD370s as its focal plane array which have been adapted from the CCD270s (CCD-Bruyeres) used on the optical imaging PLATO mission to maximise their efficiency for the low flux, low energy signals~\cite{EndicottPLATO}. The improvement in Charge Transfer Inefficiency (CTI) has been achieved by the inclusion of a supplementary buried channel (SBC) which reduces the number of traps encountered by shrinking the signals charge cloud into a smaller volume whilst undergoing a parallel transfer~\cite{ClarkeEuclidModelling}. Further detail on the function of a SBC can be found in~\cite{SeabrokeGAIA}. The responsivity has been increased from 2 to 7~$\mu$V/e by boosting the gain of the output amplifier to improve the signal to noise ratio which enables the smallest signals to be detected above the noise floor. 
 
The work presented in this paper compares the predicted parallel CTI performance improvement of the SMILE SXI CCD370 from the PLATO CCD270s against the measured performance improvement, i.e. the impact of the parallel supplementary buried channel (SBC). The SMILE SXI devices are expected to have enhanced performance when detecting soft X-rays due to the inclusion of the SBC which reduces the number of traps encountered by a small charge packet and higher gain on the output amplifier which reduces readout noise.

The performance improvements discussed compare parallel CTI predictions previously made through the Open University's (OU) bespoke code and Silvaco usage against recent lab-based trap density measurements to give an indication of the improvement in parallel CTI due to the SBC in practice compared to the pre-production estimates. The trap density datasets, provided to the OU courtesy of ESTEC, were taken on a CCD270 in 2017 as part of a Plato CCD characterisation campaign and on a CCD370 in 2019 as part of a SMILE CCD characterisation campaign. 

\section{Devices Under Test}

The two devices tested are the PLATO CCD270 and the SMILE CCD370, both are back illuminated 4510 x 4510 pixel CCDs with SiC packages, the SMILE device being a soft X-ray optimised version of the Plato CCD, specifications are illustrated in Table~\ref{Tab:DeviceSpecs}.

\begin{table}[htbp]
	\centering
	\caption{\label{Tab:DeviceSpecs} CCD270 and CCD370 design specifications.}
	\smallskip
	\begin{tabular}{|c|c|c|}
		\hline
		Specification                                      & \multicolumn{2}{|c|}{Values} \\
		\hline
		Device                                          & CCD270                           & CCD370               \\
		Manufacturing Company                           & Teledyne-e2v                     & Teledyne-e2v          \\
		Silicon Substrate                               & SiC                              & SiC           \\
		Image area size                                 & 2255 rows x 4510 columns         & 3791 rows x 4510 columns \\
		Store area size                                 & 2255 rows x 4510 columns         & 719 rows x 4510 columns  \\
		Native Pixel Size                               & 18 x 18 $\mu$m x $\mu$m          & 18 x 18 $\mu$m x $\mu$m \\
		Sensitive Silicon Thickness                     & 16 $\mu$m                        & 16 $\mu$m \\  
		Serial registers                                & 1                                & 1 \\ 
		Noise (4 MHz)                                   & 20 e $^{-}$ rms                  & 13 e $^{-}$ rms \\ 
		Dark Signal (203 K)                             & 0.4 e$^{-}$/pix/s                & 0.5 e$^{-}$/pix/s \\ 
		Output Nodes                                    & 2                                & 2 \\
		Full Well Capacity                              & 900 ke$^{-}$                     & 850 ke$^{-}$ \\ 
		SBC Capacity                                    & N/A                              & 20 ke$^{-}$ \\ 
		Output amplifier Responsivity                   & 2.2 $\mu$V/e$^{-}$               & 7 $\mu$V/e$^{-}$ \\
		\hline     
	\end{tabular}
\end{table}
Both devices have a store region with the CCD370s designed for 6x binning however all testing has been done in native pixel full frame mode.

The SBC in the CCD370 has heritage from Gaia~\cite{SeabrokeGAIA} and operates in similar way, it aims to be large enough to easily contain the soft X-rays in the SMILE science band so that the traps they interact with during a parallel transfer are minimised and hence the parallel CTI improved.   

\section{Simulated CTI Performance}
\label{sec:Simulated Performance}
In order to predict the parallel CTI performance characteristics of the PLATO CCD270 and subsequently the SMILE CCD370 a combination of experimental data and Monte Carlo simulation are used. The in-depth details of the technique are discussed in~\cite{HallNarrowChannel} however in simplified terms the pixel to pixel transfer of a range of signal sizes are simulated in Silvaco for both the standard CCD270 and a CCD270 with a SBC (as an approximation of the CCD370). This outputs a measure of the likelihood that for a given pixel geometry and signal size charge will be trapped by the dominant trap species, this is known as the "summed probability of capture" and can be subsequently interpolated across all signal levels. The dominant species is the trap with an emission time constant closest to the pumping phase delay times, which have been selected as they correspond to the expected readout speeds during the SMILE mission (at time of data being taken). At 153~K the Divacancy has a time constant of approximately 5x10$^{-4}$~s ~\cite{ParsonsCRYO} which makes it the most efficiently pumped when using the conditions specified in Table~\ref{Tab:TrapPumpingCond} and therefore the dominant trap species.    

\begin{figure}[htbp]
	\centering 
	\includegraphics[width=0.9\textwidth]{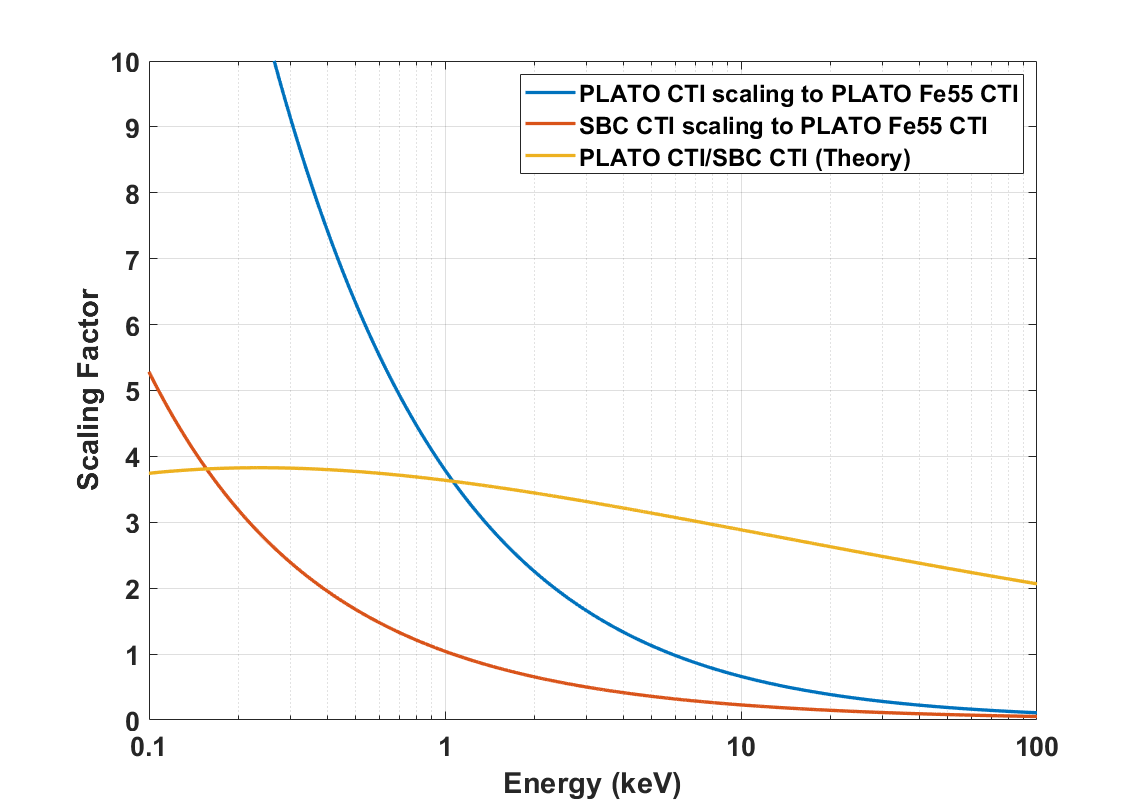}
	\caption{\label{fig:Predicted}Simulated CTI scaling for PLATO CCD270 and SMILE-like CCDs (CCD270 with a SBC).}
\end{figure}

Combining this with an 5.9~keV Fe$^{55}$ X-ray parallel CTI measurement from a CCD270 enables a CTI scaling factor to be calculated, the Plato CTI scaling shown by the blue line in Figure~\ref{fig:Predicted} across 0.1 – 100~keV is extrapolated in this way and the predicted CTI scaling for a SMILE-like device is shown by the orange line. The theoretical performance improvement of including a SBC, shown by the yellow line, can then be calculated by dividing the Plato CTI scaling by the SMILE-like scaling. The result shows that in the energy range of interest (0.2 - 2~keV) the presence of the SBC is expected to improve performance relative to a PLATO CCD270 by a factor of between three and four times.     

\section{Measured CTI Performance}

As the number of traps seen by a charge cloud as it is transferred through a CCD is directly proportional to the measured CTI~\cite{MurrayTrapPumping}, trap pumping has been used to count the traps within a CCD270 and a CCD370 to enable a comparison of the two and hence potentially validate the predicted performance improvement. As the signal increases the charge cloud occupies more of a pixels buried channel (and/or SBC) and as it grows the number of traps that capture charge also increases due to their position in the channel relative to the cloud, in this paper these are referred to as "Effective Traps". 

For any given trap species, the number of traps present will be directly proportional to the number of electrons captured in the case where the signal size is much larger than the number of traps (i.e. the capture of an electron has a negligible impact on the number of electrons in the signal packet and therefore a negligible impact on the electron density and consequently a negligible impact on the probability of capture). For signal sizes of 100-10,000 e$^-$ and the trap densities observed of <0.03 per pixel, this condition is met and the number of captured electrons (and therefore the CTI) will scale linearly with the number of traps.

If the number of traps present is high compared to the signal level such that the removal of one or more electrons from the charge packet has a significant impact on the electron density, then the probability of capture will decrease as the number of traps increases such that the relationship between CTI and the number of traps will no longer be linear. However, we are far from this regime for the measurements in this paper (high signal, low numbers of traps).

Trap pumping has become an established technique over the last 10 years~\cite{PumpingTrapProperties}~\cite{PumpingPchannel}~\cite{BushEMCCD} ~\cite{HallSiliconDI}~\cite{Woodpchannel}~\cite{SkottfeltEuclid}~\cite{Mostek} and works on the premise that as each trap species has a specific range of emission time constants individual types can be identified within a CCD by using specialised clocking schemes. These schemes, rather than moving charge packets sequentially through the device pixel to pixel instead move the charge back and forth over the same pixels (known as pumping). If the clocking is done at speeds similar to the emission time constant of a specific species then pairs of bright and dark adjacent pixels are formed as the charge is removed from one and deposited in the other, these are known as dipoles and indicate the presence of a trap. By counting the number of these traps across a range of signal levels (provided through charge injection) it is possible to plot the effective trap density against signal size. The raw data is binned to create a heat-map and the mean of each bin taken to generate the mean trap density. 

\begin{table}[htbp]
	\centering
	\caption{\label{Tab:TrapPumpingCond} Trap pumping test conditions.}
	\smallskip
	\begin{tabular}{|c|c|}
		\hline
		Variable & Value\\
		\hline
		CCD270 Dropping Phase Delay Time ($\mu$s) & 22.5 \\
		CCD270 Raising Phase Delay Time ($\mu$s) & 0 \\
		CCD370 Dropping Phase Delay Time ($\mu$s) & 14 \\
		CCD370 Raising Phase Delay Time ($\mu$s) & 1 \\
		Pump Direction & Parallel \\
		Temperature (K) & 163 +/- 1 \\
		Dipole Detection Threshold ($\sigma$)& 3 \\
		Signal Size (e$^-$) & 10$^2$-10$^5$ \\
		\hline
	\end{tabular}
\end{table} 

\begin{figure}[H]
	\centering 
	\includegraphics[width=0.85\textwidth]{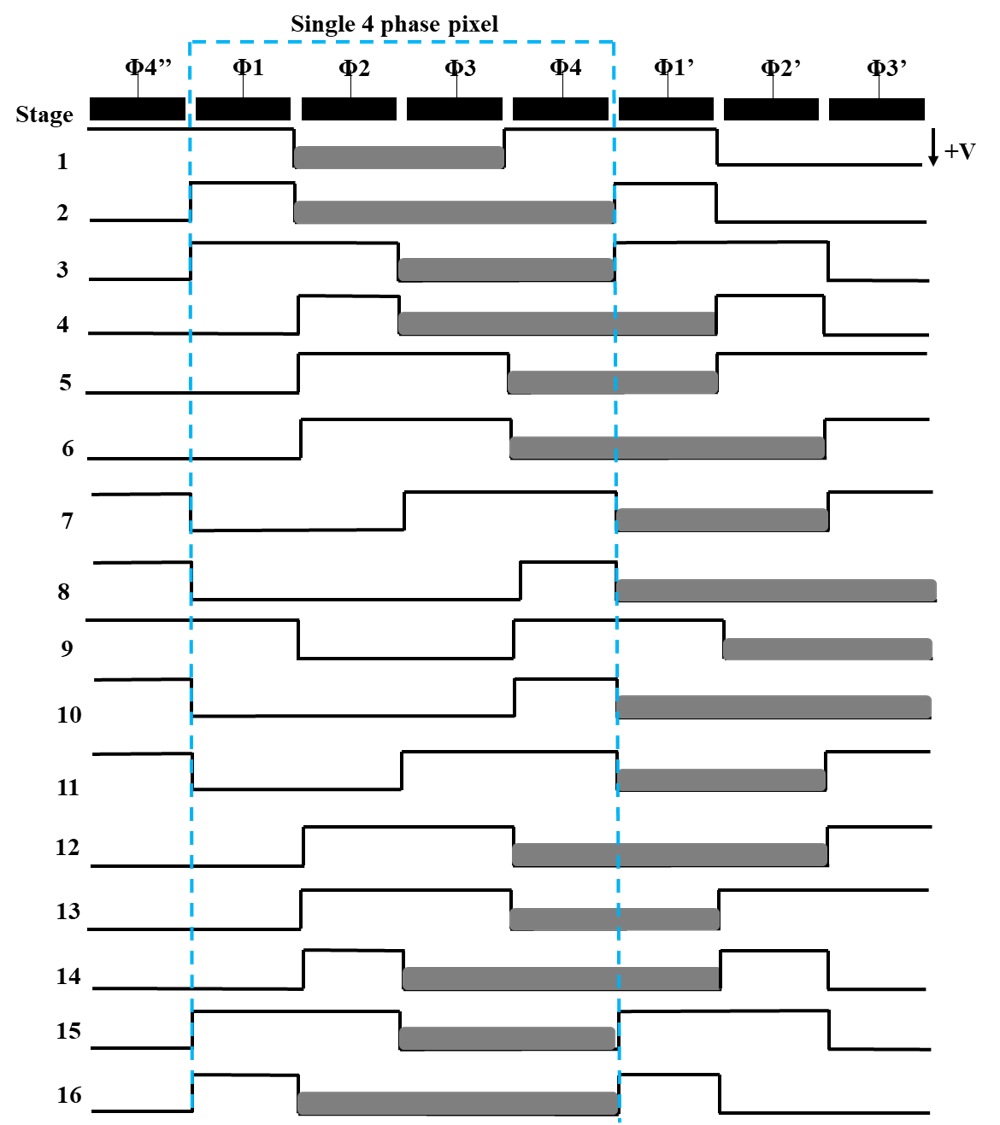}
	\caption{\label{fig:ClockingScheme} The 4-phase device trap pumping clocking scheme used.}
\end{figure}

The test conditions used during the trap pumping are summarised in Table~\ref{Tab:TrapPumpingCond}. The clocking scheme transfers the charge from under $\phi$2 and $\phi$3 to under $\phi$2 and $\phi$3 of the adjacent pixel and then back again, the clocking schematic in Figure~\ref{fig:ClockingScheme} illustrates the process. There is a slight difference in the clocking between the two devices; in the CCD370 there is a 1~$\mu$s delay after every step in which a new phase is raised and a 14~$\mu$s delay after each step in which a phase is dropped, in the CCD270 there is no delay after every step in which a new phase is raised and a 22.5~$\mu$s delay after each step in which a phase is dropped. 

The difference in both the dropping and raising phase delay time will not have a significant effect on the number of traps detected between the two clocking schemes, and hence the data from the CCD270 and CCD370 can be explicitly compared. The effect will be insignificant as the method for trap counting  relies upon a threshold signal limit (3 $\sigma$) above the mean column signal during the data analysis pipeline. As the rising and falling delay times are slightly different, traps will pump with a slightly different efficiency leading to a change in the high/low signal of each dipole. However, the small change in delay time (maximum 8.5~$\mu$s) will mean that the number of traps dropped below the detection threshold will be negligible.

\begin{figure}[htbp]
	\centering 
	\includegraphics[width=0.70\textwidth]{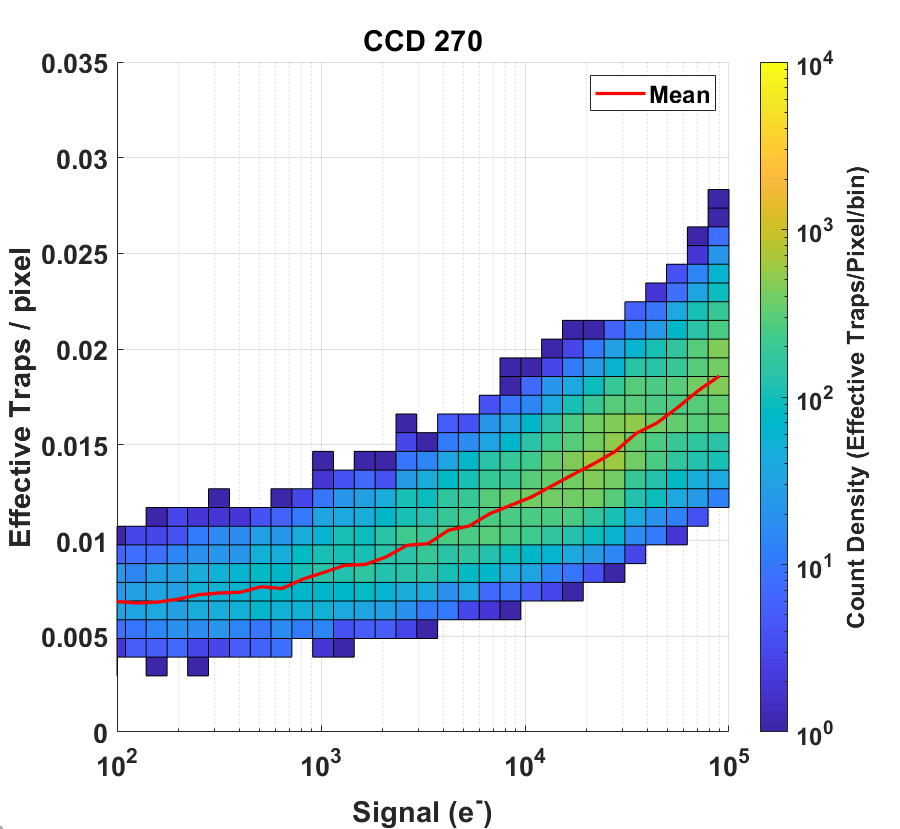}
	\caption{\label{fig:270density} PLATO CCD270 trap density.}
\end{figure}

\begin{figure}[htbp]
	\centering 
	\includegraphics[width=0.70\textwidth]{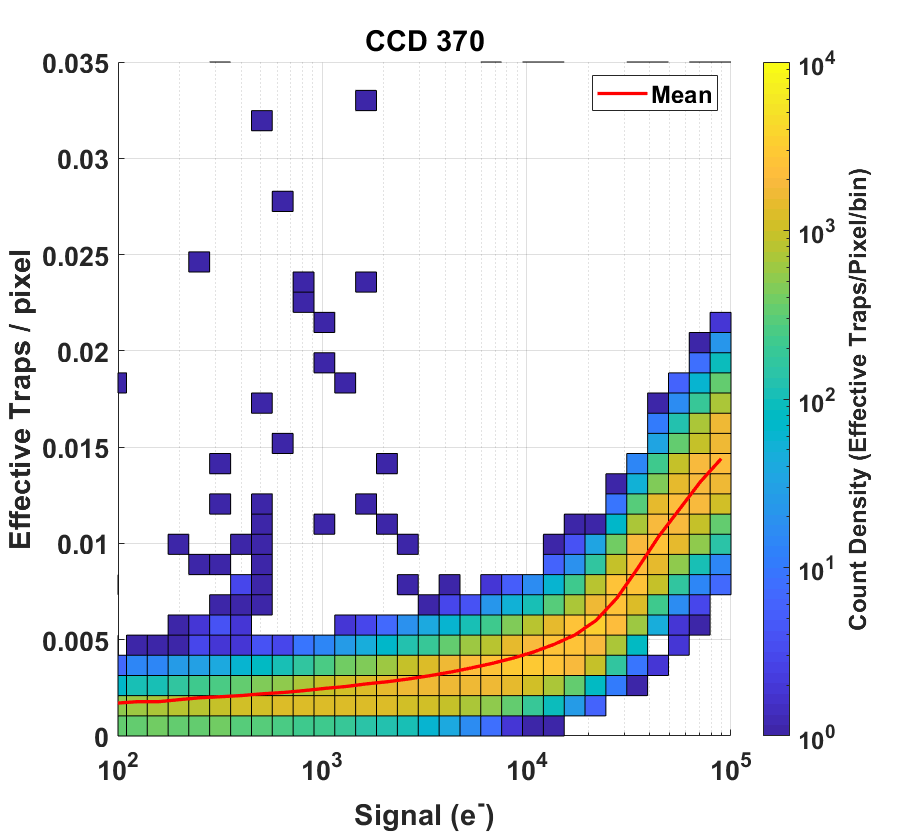}
	\caption{\label{fig:370density} SMILE CCD370 trap density.}
\end{figure}

The effective trap density for the CCD270 as a function of signal size is in Figure~\ref{fig:270density} and for the CCD370 in Figure~\ref{fig:370density}. As expected the density increases with signal size, the capacity of the SBC can also be seen in Figure~\ref{fig:370density} at around 20 ke$^-$ where the effective trap density suddenly increases as the charge cloud completely fills the SBC and starts to expand over the buried channel, this capacity is in line with the design specification in Table~\ref{Tab:DeviceSpecs}.  

\begin{figure}[H]
	\centering 
	\includegraphics[width=0.8\textwidth]{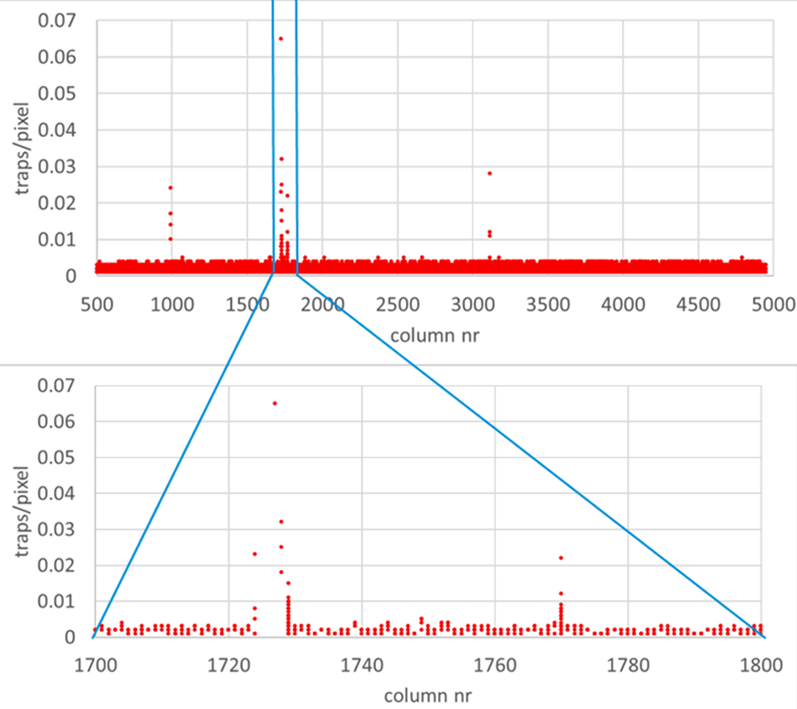}
	\caption{\label{fig:370Traps_Col} Trap density across the CCD370 columns.}
\end{figure}

The high trap density outliers in Figure~\ref{fig:370density} are only found in specific columns, this is illustrated by plotting trap density against column number as in Figure~\ref{fig:370Traps_Col}. These columns correlate with known partially dead columns within the CCD and are thought to produce spurious dipoles in the trap pumping analysis algorithm. However, the outliers are orders of magnitude lower than the peak counts and do not impact the mean in a meaningful way. 

\section{Comparison of Simulation to Experimental Results}

To find the experimental performance improvement between the PLATO and SMILE devices the mean effective trap density found in the CCD270 is divided by the CCD370 values, as in Equation~\ref{eq :FactorImprovement} where $T\rho_{270}$ = mean CCD270 effective trap density and $T\rho_{370}$ = mean CCD370 effective trap density.

\begin{equation}\label{eq :FactorImprovement}
Performance~Improvement = \frac{T\rho_{270}}{T\rho_{370}} 
\end{equation}

\begin{figure}[htbp]
	\centering 
	\includegraphics[width=0.9\textwidth]{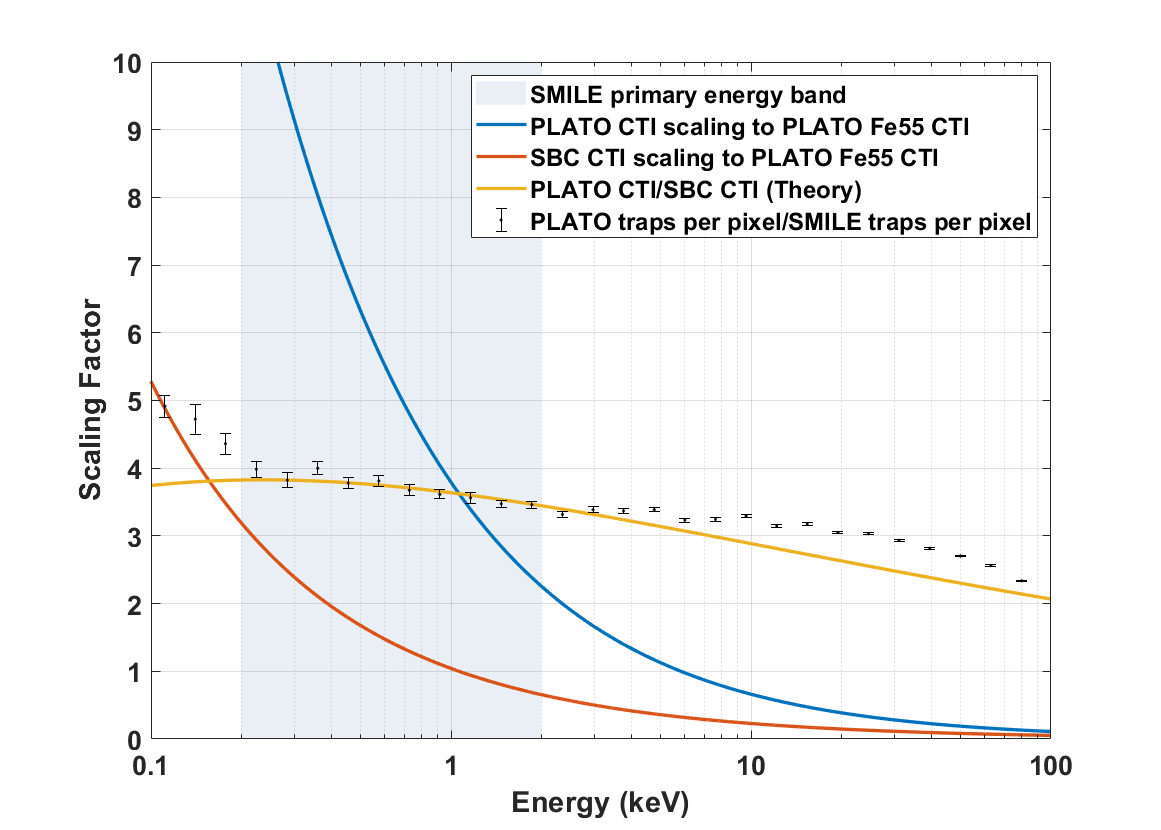}
	\caption{\label{fig:Measured} Simulated CTI scaling for Plato and SMILE-like CCDs overlaid with measured CCD270/CCD370 effective trap density data.}
\end{figure}

Overlaying the measured performance improvement on the predicted values from Section~\ref{sec:Simulated Performance} in Figure~\ref{fig:Measured} shows that they agree very well, with particularly good agreement within the 0.2 – 2~keV key energy range for the SMILE SXI. The predicted improvement factor appears to underestimate the measured values slightly outside of the SXI energy range. At the low energies this is thought to be due to a combination of the sampling resolution of the Silvaco-generated electron densities used during the simulation and the threshold setting used to extract the dipoles. The difference in the prediction above 5~keV is likely due to differences in the measured SBC capacity compared to the simulations. 

The uncertainty on the simulated and measured data will also be larger than displayed as it does not account for variations in device fabrication batches, which potentially have different numbers of native traps which would therefore directly affect the scaling factor. Slightly different measurement temperatures and clock phase times between the devices are also not taken into account, although their impact is expected to be small.

\section{Summary}

The simulation of the predicted CTI performance is built upon previous work ~\cite{HallNarrowChannel} which looked at modelling charge transport in simpler devices that did not include a SBC. The SMILE SXI CCDs have been optimised for soft X-rays and as part of that process a SBC has been embedded into the parallel buried channel to entirely contain the signals charge clouds and reduce the number of traps it encounters as it is transported through the CCD. The simulation predicts a factor of three to four times improvement in parallel CTI performance over the key SMILE SXI energy range (0.2-2~keV) with this gradually reducing to a factor of 2 as the X-ray energy approaches 100~keV. 

Trap pumping has been carried out on both a Plato CCD270 and a SMILE CCD370 and this data has been used to count the number of effective traps per pixel present in each device as a function of signal size. As the number of traps the charge cloud interacts with is directly proportional to the CTI performance these trap densities have then be used to calculate the performance improvement factor between the two device types.   

A comparison of the simulated to experimental results show good agreement, especially within the SMILE SXI science band. Therefore, the optimisation made to mitigate radiation induced charge transfer inefficiency in the design of the SMILE CCD370 has now been shown to perform as expected, thus adding confidence to one key aspect of the early performance modelling of the SMILE SXI instrument. 

Beyond the SMILE mission, the accuracy of the performance improvement prediction with a more advanced geometry is an indication that the Silvaco and Matlab simulations are good models of real devices and are suitable to be used in the future to help understand a devices behaviour and potentially inform the design process pre-fabrication. It also adds confidence to the effectiveness of using an SBC, especially in missions with small signals.


\begin{thebibliography}{99}

\bibitem{RaabSMILE}
W. Raab et al. \emph{SMILE: A joint ESA/CAS mission to investigate the interaction between the solar wind and Earth’s magnetosphere}, \emph{Proc. SPIE 9905}, {\bf Article Number} 990502., 2016

\bibitem{DennerlChargeExchange}
K. Dennerl et al. \emph{Solar system X-rays from charge exchange processes}, \emph{Astron. Nachr. 333}, {\bf Article Number} 324, 2012

\bibitem{WalshSXIscience}
B. Walsh et al. \emph{Wide field-of-view soft X-ray imaging for solar wind magnetosphere interactions}, \emph{J. Geophys. Res. 121}, {\bf Article Number} 3353., 2016

\bibitem{HollandDamage}
A. Holland et al. \emph{Techniques for minimizing space proton damage in scientific charge coupled devices}, \emph{IEEE Trans. Nucl. Sci. 38}, {\bf Article Number} 1663, 1991

\bibitem{EndicottPLATO}
J. Endicott et al. \emph{Charge-coupled devices for the ESA PLATO M-class Mission}, \emph{Proc. SPIE 8453}, {\bf Article Number} 84531J, 2012

\bibitem{ClarkeEuclidModelling}
A. Clarke et al. \emph{Modelling charge storage in Euclid CCD structures}, \emph{JINST 7}, {\bf Article Number} C01058, 2012

\bibitem{HallNarrowChannel}
Hall, D. J.; Skottfelt, J.; Soman, M. R.; Bush, N. and Holland, A. \emph{Improving radiation hardness in space-based Charge-Coupled Devices through the narrowing of the charge transfer channel}, \emph{Journal of Instrumentation, 12}, {\bf Article Number} C12021, 2017

\bibitem{ParsonsCRYO}
Parsons, S.; Buggey, T.; Holland, A.; Sembay, S.; Randall, G.; Hetherington, O.; Yeoman, D.; Hall, D.; Verhoeve, P. and Soman, M. \emph{Effects of temperature anneal cycling on a cryogenically proton irradiated CCD}, \emph{Journal of Instrumentation, 16(11)}, {\bf Article Number} P11005, 2021.

\bibitem{MurrayTrapPumping}
Murray, N. Burt, D. Hall, D. and Holland, A. \emph{The relationship between pumped traps and signal loss in buried channel CCDs}, \emph{UV/Optical/IR Space Telescopes and Instruments: Innovative Technologies and Concepts VI, SPIE}, {\bf Article Number} 8860 0H, 2013

\bibitem{SeabrokeGAIA}
Seabroke, G. M. and Prod'homme, T. and Murray, N. J. and Crowley, C. and Hopkinson, G. and Brown, A. G. A. and Kohley, R. and Holland, A. \emph{Digging supplementary buried channels: investigating the notch architecture within the CCD pixels on ESA's Gaia satellite},\emph{Monthly Notices of the Royal Astronomical Society}, {\bf Volume} 430, 2013,\emph{10.1093/mnras/stt121}

\bibitem{PumpingTrapProperties}
Hall, D.; Murray, N.; Holland, A.; Gow, J.; Clarke, A. and Burt, D.  \emph{Determination of in situ trap properties in CCDs using a "single-trap pumping" technique.} \emph{IEEE Transactions on Nuclear Science, 61(4) pp.} 1826–1833. (2014)

\bibitem{PumpingPchannel}
Wood, D.; Hall, D.; Murray, N.; Gow, J. and Holland, A. \emph{Studying charge-trapping defects within the silicon lattice of a p-channel CCD using a single-trap “pumping” technique.} \emph{Journal of Instrumentation, 9, article no.} C12028. (2014)

\bibitem{BushEMCCD}
Bush, N.; Hall, D.; Holland, A.; Burgon, R.; Murray, N.; Gow, J.; Jordan, D.; Demers, R.; Harding, L.K.; Nemati, B.; Hoenk, M.; Michaels, D. and Peddada, P. (2016). \emph{Cryogenic irradiation of an EMCCD for the WFIRST coronagraph: preliminary performance analysis.} \emph{High Energy, Optical, and Infrared Detectors for Astronomy VII, Society of Photo-Optical Instrumentation Engineers (SPIE)}, article no. 99150A.

\bibitem{HallSiliconDI}
Hall, D.; Wood, D.; Murray, N.; Gow, J.; Chroneos, A. and Holland, A. (2017). \emph{In situ trap properties in CCDs: the donor level of the silicon divacancy.} \emph{Journal of Instrumentation, 12}, article no. P01025.

\bibitem{Woodpchannel}
Wood, D.; Hall, D.; Gow, J.; Skottfelt, J.; Murray, N.; Stefanov, K. and Holland, A. (2017). \emph{Evolution and impact of defects in a p-channel CCD after cryogenic proton-irradiation.} \emph{IEEE Transactions on Nuclear Science, 64(11)} pp. 2814–2821.

\bibitem{SkottfeltEuclid}
Skottfelt, J.; Hall, D.J.; Dryer, B.; Bush, N.; Campa, J.; Gow, J.P.D.; Holland, A.D.; Jordan, D. and Burt, D. (2017). \emph{Trap pumping schemes for the Euclid CCD273 detector: characterisation of electrodes and defects.} \emph{Journal of Instrumentation, 12}, article no. C12033.

\bibitem{Mostek}
N. J. Mostek, C. J. Bebek, A. Karcher, W. F. Kolbe, N. A. Roe and J. Thacker. \emph{Charge trap identification for proton-irradiated p+ channel CCDs}, \emph{Proc. SPIE 7742}, 2010.

\end{thebibliography}
\end{document}